\documentclass[aps,prd,onecolumn,groupedaddress,nofootinbib,preprint,11pt,a4paper]{revtex4-1}
\usepackage[colorlinks=true,citecolor=magenta,linkcolor=blue,breaklinks=true]{hyperref}
\usepackage{amsmath}
\usepackage{graphicx,graphics}
\usepackage{amssymb}
\usepackage{xcolor}
\usepackage{caption}
\usepackage[caption=false,labelformat=simple]{subfig}

\usepackage{booktabs}
\usepackage{cellspace}
\newcommand{\tanb}{\tan\beta}
\newcommand{\lamhAA}{\lambda_{hAA}}

\begin{document}
\title{Improving probes of $hAA$ coupling in the Type-X two Higgs
Doublet Model scenario: the crucial role of $\tau$-jet charge identification}

\author{Biswarup Mukhopadhyaya}
\email{biswarup@iiserkol.ac.in}

\author{Sirshendu Samanta}
\email{ss21rs027@iiserkol.ac.in}

\author{Tousik Samui}
\email{tousik.pdf@iiserkol.ac.in}

\author{Ritesh K. Singh}
\email{ritesh.singh@iiserkol.ac.in}
\affiliation{Department of Physical Sciences, \\ Indian Institute of
Science Education and Research Kolkata, \\ Mohanpur, 741246, India.}

\begin{abstract}
The exotic decay modes of the already discovered 125-GeV scalar into a pair of light pseudoscalars is a good probe of those new physics scenarios where such pseudoscalars exist. Searches in the mass region where the pseudoscalar ($A$) is lighter than 62.5 GeV have yielded null findings so far. No search has yet examined $m_A > 62.5$ GeV where the cross section is suppressed by the off-shell pseudoscalar. We point out a possible enhancement of the sensitivity of probing $hAA$ coupling in the context of a Type-X two Higgs doublet model. We focus on $h\to A A^{(*)} \to 4\tau$, and select events with two same-sign $\tau$-jets along with a pair of same-sign leptons. This enables much more effective background elimination than in the erstwhile proposed channels. Taking two values of the $\tau\tau A$ coupling into account, we obtain limits on the $hAA$ coupling that can be probed at 2$\sigma$ and 3$\sigma$ significance, for $m_A$ ranging up to 85 GeV. For $m_A < 62.5$ GeV, too, we find the probe through our suggested channel exhibits considerable improvement upon the usual $2\mu 2\tau$-based searches conducted at the LHC. Within this region, achieving the reach of $\lamhAA$ coupling at 3000 fb$^{-1}$ luminosity using our strategy would require approximately $\sim$\,3.6$\times 10^{5}$ fb$^{-1}$ luminosity using conventional $2\mu 2\tau$-based searches.
\end{abstract}

\maketitle


\section{Introduction}
Although the Standard Model (SM) hypothesizes a single electroweak symmetry breaking $SU(2)$ doublet, observed data allow the participation of other scalar multiplets\,\cite{Branco:2011iw,Bhattacharyya:2015nca,Aggleton:2016tdd,Davoudiasl:2022mav,Ivanov:2017dad,Robens:2021lov}. Furthermore, the observed repetition of fermion families provides a spur to the idea of an extended electroweak symmetry-breaking sector. The first stride in this direction leads to two Higgs doublet models (2HDM), of which several types exist\,\cite{Barger:1989fj,Grossman:1994jb,Aoki:2009ha,Branco:2011iw,Wang:2022yhm}. These various types essentially differ from each other in the Yukawa interactions of the two doublets. The phenomenological constraints on the masses of the various
spinless physical states depend on these Yukawa couplings since they affect production as well as decay channels, and also the role of these scalars in loop-induced rare processes. A question naturally arising, and of phenomenological interest, is: can the masses of some spin-zero particles\,\cite{Cao:2009as,Chun:2017yob,Chun:2018vsn,Dey:2021alu} in certain kinds of 2HDM lie well below the 125-GeV scalar\,\cite{ATLAS:2012yve,CMS:2012qbp} that has already been discovered?

The physical pseudoscalar state $A$ in 2HDM is the readiest candidate here. This is because this state is more likely to be well-separated in mass from the heavier neutral scalar and the charged scalar, which are more constrained, in an inter-related manner, from collider searches\,\cite{Han:2022juu}, rare decays\,\cite{Baum:2007ts,T_Barakat_1998,Misiak:2017bgg}, and also electroweak precision measurement data\,\cite{Mukhopadhyaya:2023akv,Lu:2022bgw}. At the same time, a light pseudoscalar is phenomenologically viable when its coupling with quarks is suppressed\,\cite{Dey:2021alu,Dey:2021pyn}. Out of the major 2HDM types, this is possible in the Type-X scenario, where one of the doublets couples to quarks only, and the other, only to leptons, engineered by an appropriate $\mathbb{Z}_2$-symmetry in the Yukawa sector. 

The neutral pseudoscalar $A$ in Type-X 2HDM has unsuppressed couplings with leptons only if the so-called  `alignment limit' has to be adhered to\,\cite{Gunion:2002zf,Bhattacharyya:2015nca,Karmakar:2018scg}. It is of obvious interest to look for it at the Large Hadron Collider (LHC)\,\cite{Lyndon:Evans2008,Hou:2022nyh,Adhikary:2022jfp}. Moreover, it is useful to find out about the parameters of the scalar potential in such a scenario using data from the high-luminosity LHC (HL-LHC)\,\cite{Contardo:2020886,CERN-LHCC-2020-004,Ma:2024deu}. For instance, the pseudoscalar pair produced in the decays of $h$, the 125-GeV scalar, is driven by the $hAA$ interaction strength. In this work, we suggest new ways of extracting such information, using some features of the $4\tau$ final state arising from $h\rightarrow AA$.

For the case where $h\rightarrow AA$ is kinematically allowed, searches have been performed in the $4\mu$\,\cite{ATLAS:2018coo,CMS:2018jid}, $4b$\,\cite{ATLAS:2018pvw,ATLAS:2020ahi}, $2b2\tau$\,\cite{CMS:2018zvv}, $2\mu 2\tau$ and $4\tau$\,\cite{ATLAS:2015unc,CMS:2017dmg,CMS:2018qvj,CMS:2019spf,CMS:2020ffa} channels at the LHC. These searches tend to disfavour a pseudoscalar lighter than $m_h /2$.  
Upper limits on the $hAA$ interaction strength (depending, of course, on the other 2HDM parameters) are also obtainable. However, these limits pertain to on-shell production of both the $A$'s in $h$-decay. With one $A$ becoming off-shell, the usual search channels get overridden by backgrounds, making studies at the HL-LHC difficult.

Two questions can be asked now: (a) Can the light pseudoscalar scenario be probed beyond $m_A \ge m_{h}/2$? And (b) What are the ranges of the $hAA$ interaction strength, which can be probed for various values of $m_A$, thus providing insight into the
2HDM scalar potential? We address both of these questions here and arrive at positive answers. 

The special features of this work lie in the following points:
\begin{itemize}
\item We suggest looking for $h\rightarrow AA^{(*)} \rightarrow 4\tau$ yielding final states where two same-sign $\tau$'s decay leptonically, while the two remaining ones (again of the same sign) undergo hadronic decay in one-prong and three-prong modes. One can then concentrate on signals consisting of two same-sign leptons plus a pair of same-sign $\tau$-jets. Given the recent estimates of 
$\tau$-jet charge identification efficiency\,\cite{CMS:2022prd},  we show such signals, with appropriate cuts that we propose, to be significantly more effective in reducing backgrounds\,\cite{Mukhopadhyaya:2023akv}. Thus, our event selection criteria make it possible to extend the probe of the $hAA$ coupling strength well beyond the reach of strategies adopted earlier\,\cite{ATLAS:2015unc,CMS:2017dmg,CMS:2018qvj,CMS:2019spf,CMS:2020ffa}.
\item The level at which the $hAA$ interaction strength can be probed using the aforementioned signal depends also on the $A\tau\tau$ Yukawa interaction strength since one is thinking here in terms of an off-shell $A$ driving our signal. We thus benchmark our probes in terms of the 
$A\tau\tau$ coupling strength. This gives further insight into the parameter space of the theory.
\end{itemize}

Incidentally, a low-mass pseudoscalar in Type-X 2HDM scenario has been often suggested as a solution\,\cite{Wang:2014sda,Ilisie:2015tra,Abe:2015oca,Ghosh:2020tfq,Ghosh:2021jeg,Jueid:2021avn,Dey:2021pyn,Iguro:2023tbk} to the deficit in $(g-2)_\mu$\,\cite{Muong-2:2023cdq}, the muon anomalous magnetic moment, predicted in the standard model (SM)\,\cite{Aoyama:2020ynm}. However, claims from lattice estimates\,\cite{FermilabLattice:2017wgj,Giusti:2018mdh,Giusti:2019xct,Shintani:2019wai,FermilabLattice:2019ugu,Gerardin:2019rua,Blum:2019ugy,Borsanyi:2020mff,Ce:2022kxy,ExtendedTwistedMass:2022jpw,Chao:2022ycy} of 
$(g-2)_\mu$, as well as mutually contradictory data-driven estimates\,\cite{Davier:2019can,Keshavarzi:2019abf,CMD-3:2023alj,Muong-2:2023cdq} of the hadronic vacuum polarisation contributions, make the scenario more fluid on this front. Our discussion in this paper is therefore dissociated from the $(g-2)_\mu$ issue, and we treat the search for a light pseudoscalar as important by its own right.

The organization of the paper is as follows: a brief description of
the Type-X 2HDM and parameter space limitations are discussed in
Section \ref{sec:model}, which serves as the justification for using
benchmark points. The methodology of our analysis has been discussed
in Section \ref{sec:methodology}, and the suggested signal, including
the discussion of its backgrounds, is presented in Section
\ref{sec:sig-bkg}. The results are then presented and examined in
Section \ref{sec:result-dicuss}. We then summarize our work in Section
\ref{sec:summ}.

\section{Type-X 2HDM: parameters and constraints} \label{sec:model}
The general framework of 2HDM extends the SM scalar sector by another
$SU(2)_L$ Higgs doublet with hypercharge same as the SM Higgs doublet.
After the electroweak symmetry breaking (EWSB), the two doublets
$\Phi_1$ and $\Phi_2$ receive their vacuum expectation values (vev)
$v_1$ and $v_2$, respectively. These two vevs are usually
re-parametrized in terms of $\tanb \equiv v_2/v_1$ and
$v \equiv \sqrt{v_1^2 + v_2^2}$.
As it involves two doublets, it is natural to have flavour-changing
neutral current (FCNC)\,\cite{Sher:2022aaa}, for which no significant
experimental evidence has been observed so far, to occur at the tree
level after the EWSB. To avoid the FCNC, an extra symmetry, namely
$\mathbb{Z}_2$, is usually imposed, and odd charges are assigned to
$\Phi_1$ and a set of right-handed fermions. Depending on this set,
four different types of 2HDM occur. These four types are Type-I, Type-II, Type-X, and Type-Y (also called flipped type). A comprehensive discussion of these four types, including their
phenomenological consequences, can be found in Ref.\,\cite{Branco:2011iw,Wang:2022yhm}.
With this $\mathbb{Z}_2$ symmetry, the $CP$-conserving scalar
potential can be written as:
\begin{eqnarray}
V_\text{scalar}
&=&  m_{11}^2 \Phi_1^\dagger \Phi_1
    + m_{22}^2 \Phi_2^\dagger \Phi_2
    + \lambda_1 \left(\Phi_1^\dagger \Phi_1\right)^2
    + \lambda_2 \left(\Phi_2^\dagger \Phi_2\right)^2 + \lambda_3 \left(\Phi_1^\dagger \Phi_1\right)
     \left(\Phi_2^\dagger \Phi_2\right) \nonumber \\
& & + \lambda_4 \left(\Phi_1^\dagger \Phi_2\right)
     \left(\Phi_2^\dagger \Phi_1\right) + \left\{ -m_{12}^2 \Phi_1^\dagger \Phi_2
    + \frac{\lambda_5}{2} \left(\Phi_1^\dagger \Phi_2\right)^2
    + \text{h.c.} \right\}, \label{eqn:LagS}
\end{eqnarray}
Note that, although the quadratic term $m_{12}^2 \Phi_1^\dagger\Phi_2$
violates $\mathbb{Z}_2$ symmetry softly, it does not induce any further
FCNC\,\cite{Arco:2022jrt}. After the spontaneous breaking of the electroweak symmetry,
three degrees of freedom are absorbed by the electroweak gauge bosons
to receive their masses. The remaining modes appear as physical
scalars as one pseudoscalar $A$, one charged scalar $H^\pm$, and two
neutral $CP$-even scalars $H$ and $h$. The mass matrices of the
charged scalars and pseudoscalars are diagonalized by the same mixing
angle $\beta = \tan^{-1}(v_2/v_1)$, while the diagonalization of the
neutral CP-even scalar mass matrix requires a different angle, say,
$\alpha$. For our study, we concentrate on a paradigm where $h$ is
identified as the observed 125-GeV scalar, the pseudoscalar is lighter
than $h$, and other scalars are heavier than these two.
This parameter region can be achieved by appropriately adjusting the
six free parameters, namely $\lambda_1$, $m_H$, $m_{H^\pm}$, $m_A$,
$\tan \beta$, $m_{12}^2$ after redefining the Lagrangian parameters
and setting the mass of $h$ and the vev $v$ to their respective
observed values at 125~GeV and 246~GeV.

The constraint arises from the theoretical side to ensure the scalar
potential remains bounded from below\,\cite{PhysRevD.75.035001,Chang:2015goa}, which
gives the following conditions on the $\lambda$ parameters.
$$\lambda_{1,2} > 0, \qquad \lambda_3 > -\sqrt{\lambda_1\lambda_2},
\quad \text{and} \quad \lambda_3 + \lambda_4 - |\lambda_5| > -\sqrt{\lambda_1\lambda_2}\,.$$
The second source of theoretical constraints comes from the tree-level
unitarity preservation\,\cite{Arhrib:2000is,Ginzburg:2005dt} in pure scalar-scalar
elastic and inelastic scattering processes
($S_1S_2\rightarrow S_3 S_4$) dominated by quartic interactions. From
the mode partial wave decomposition of the scattering amplitudes, this
unitarity constraint can be extracted from the condition
$\Re(a_l)^2 + \Im(a_l)^2 = |a_l|^2 = \Im(a_l)$, where $a_l$ is the
$\ell^{\rm th}$ mode of partial wave. In graphical representation, this
constraint, in turn, puts $|\Re(a_l)| < \frac{1}{2}$. These
theoretical constraints put restrictions on the parameter space,
details of which can be found in Ref.\,\cite{Arhrib:2000is}.
Furthermore, to have a perturbative Lagrangian, the quartic couplings
should obey $|\lambda_i| < 4\pi$, ($i=1,2,\cdots,5$). We have verified
that the parameter points chosen for further analysis satisfy these
constraints.

In this work, we focus on the Type-X
scenario, in which the lepton Yukawa coupling is with the doublet
$\Phi_1$, and quark Yukawa couplings are with the doublet $\Phi_2$.
With these charge assignments, the Yukawa coupling after the EWSB
becomes
\begin{eqnarray}
      \mathcal{L}_\text{Yukawa}
&=& - \sum_{f} \frac{m_f}{v} \left(\xi_f^h \bar f f h
    + \xi_f^H \bar f f H - i \xi_f^A \bar f \gamma_5 f A \right)
      \nonumber\\
& & - \sum_{u,d} \frac{\sqrt{2}}{v}\Big[V_{ud}^\text{CKM}
      \left(m_u\xi_u^A \bar{u}_R d_L
    + m_d\xi_d^A \bar{u}_L d_R\right)H^+
    + m_\ell \xi_\ell^A \bar{\nu}_L\ell_R H^+
    + \text{h.c.}\Big], \label{eq:yukawa}
\end{eqnarray}
where
\begin{eqnarray}
&& \xi_{u,d}^h = \frac{\cos\alpha}{\sin\beta}, \qquad
\xi_{u,d}^H = \frac{\sin\alpha}{\sin\beta}, \qquad
\xi_{u,d}^A = \pm \cot\beta, \label{eqn:hqq-coup} \nonumber\\
&&\xi_\ell^h = -\frac{\sin\alpha}{\cos\beta}, \qquad \xi_\ell^H = \frac{\cos\alpha}{\cos\beta}, \qquad \xi_\ell^A = \tan\beta. \label{eqn:hll-coup}
\end{eqnarray}
As a phenomenological consequence, mainly dictated by the Yukawa
coupling of the top quark, one needs to take large $\tan\beta$ in this
setup. The experimental observations, mainly the resemblance of the
125-GeV scalar to the SM Higgs boson, sets
$\beta-\alpha \simeq \pi/2$, which is the so-called `alignment limit'.
In this large $\tan\beta$ and alignment limit, Type-X 2HDM shows an
interesting scenario where new scalars, namely $H^\pm$, $A$, and $H$,
couple feebly to the quarks but strongly couple to the leptons. This
has consequences, especially in the low $m_A$ region, in generating
large new physics contributions to $(g-2)_\mu$, in which a significant
excess is seen in the BNL\,\cite{Muong-2:2006rrc} and
FNAL\,\cite{Muong-2:2023cdq} experiments. There are ongoing debates
about whether or not this excess really exists since fresh
calculations of the SM prediction indicate
otherwise\,\cite{Borsanyi:2020mff,CMD-3:2023alj}. Nevertheless, this
large coupling of $A$ to lepton plays an important role in our study
since we are also interested in the $h\to A A \to 4\tau$ channel.
This is because the Yukawa coupling in Eq.~\eqref{eq:yukawa} between
the pseudoscalar and leptons has an enhancement factor $\tan\beta$, and,
due to the highest mass of $\tau$ lepton, the dimensionless
$y_{\tau\tau A}$ coupling is dominant in the alignment limit. So, the
primary decay mode of the pseudoscalar is in the $\tau$ channel
($\sim$100\%).

One source of the constraints on the model parameters comes from the
measurement of the properties of the observed 125-GeV scalar. Although
the observed scalar mostly resembles the SM Higgs boson, it still
leaves a tiny window to keep our model parameters. We have performed a
thorough study of these constraints, primarily in the measurement of
Higgs signal strength, defined as the ratio of the production
cross section of the 125-GeV scalar in a particular channel to the SM
prediction of that. For this, we used
{\tt HiggsSignals}\,\cite{bechtle2021higgssignals} which is
implemented within the package
{\tt HiggsTools}\,\cite{bahl2023higgstools}. The second experimental
constraints come from the search for other scalars for which we do not
have a clear signal, and thus, the cross sections of the production of
these other scalars ($H^\pm$, $H$, and $A$ in our model) is measured
to be less than the detectable values.
These searches have been performed in the Large Electron-Positron
(LEP) collider  and in the LHC in the
$h \rightarrow AA$\,\cite{ATLAS:2018emt,ATLAS:2018pvw,CMS:2018qvj,CMS:2018nsh,CMS:2018zvv,CMS:2019spf},
$H \rightarrow ZZ/W^+ W^-$\,\cite{CMS:2018amk,ATLAS:2018sbw,CMS:2019bnu},
$H \rightarrow hh$\,\cite{ATLAS:2018rnh,ATLAS:2018hqk,CMS:2018ipl} etc.
channels. We have used
{\tt HiggsBounds}\,\cite{bechtle2020higgsbounds} using
{\tt HiggsTools}\,\cite{bahl2023higgstools} to verify the validity of
the model parameters at 95\% C.L.

The other set of constraints comes from the measurement of electroweak
precision observables (EWPO), which are defined through
Peskin-Takeuchi electroweak oblique parameters, $S$, $T$ and
$U$\,\cite{PhysRevD.46.381}. These observables have been measured with good precision at the LEP\,\cite{ALEPH:2005ab}. For new physics, the
values of specified parameters deviate from the standard model's
value, which is exactly zero. The parameter $U$ can be taken as zero
due to having a suppression factor compared to $S$ and $T$. We have
taken the current best-fit result, $S=-0.01\pm0.07$, and
$T=0.04\pm0.06$ in the limit $U=0$\,\cite{ParticleDataGroup:2020ssz}. All
the parameter points we consider are inside the $90\%$ C.L.~contour in
the $S-T$ plane.

\section{Collider Study}
\subsection{Analysis Methodology} \label{sec:methodology}
We are interested in the search of pseudoscalar ($A$) in $h\to A A$
channel in the gluon-gluon fusion (ggF) production of the $h$. In our
model, the $h g g$ coupling resembles the SM values in the high
$\tan\beta$ and alignment limit. So, the coupling that can be probed
in this channel is the coupling $\lamhAA$, the coupling between the
Higgs boson $h$ and the pseudoscalar $A$. In terms of the other
parameters of the theory, this coupling takes the following form in
the alignment limit:
\begin{eqnarray}
   \lamhAA &\approx& v\sin^4\!\beta\ (\lambda_3 + \lambda_4 - \lambda_5).
\end{eqnarray}
This non-zero coupling $\lamhAA$ contributes to the branching ratio
(BR) of $h$ to its non-standard decays whenever $m_A < m_h/2$. The expression for the BR in this mode can be written
as
\begin{eqnarray}
{\rm BR}(h \rightarrow AA) &=& \frac{\Gamma (h \rightarrow AA)}{\Gamma^{\rm SM}_h + \Gamma (h \rightarrow AA)}, \label{eq:brhAA}
\end{eqnarray}
where $\Gamma^{\rm SM}_h$ is the decay width of $h$ to SM
modes and the partial decay with for
the decay $h \rightarrow AA$ takes the form
\begin{eqnarray}
\Gamma(h \rightarrow AA) &=& \frac{1}{32\pi}  \frac{\lamhAA^2}{m_h}\sqrt{1 - \frac{4m_A^2}{m_h^2}}. \label{eq:dwhAA}
\end{eqnarray}
The partial decay widths of $h$ to SM modes ($\Gamma_h^{\rm SM}$) remain almost unaltered with respect to those of the SM Higgs boson. This is because the fermion coupling modifiers $\xi^h_{u, d, \ell}$ in Eq.~(\ref{eqn:hll-coup}) remain almost unity in the alignment limit. In this limit, the $hWW$ and $hZZ$ coupling modifiers, which take the form $\sin(\beta-\alpha)$, also remain close to one\,\cite{Branco:2011iw}. Therefore, for our analysis, we have used the value of $\Gamma_h^{\rm SM}$ as recommended by the LHC Higgs Cross Section Working Group (LHCHCSWG)\,\cite{LHCHiggsCrossSectionWorkingGroup:2011wcg,Denner:2011mq,Dittmaier:2012vm,LHCHiggsCrossSectionWorkingGroup:2013rie,LHCHiggsCrossSectionWorkingGroup:2016ypw}, which uses the package {\tt HBDECAY}\,\cite{Djouadi:1997yw,Djouadi:2018xqq} to calculate the widths and branching ratios for all kinematically allowed channels, incorporating higher-order corrections. The calculation considers all relevant decay channels for the Standard Model Higgs boson, including $h \rightarrow f\Bar{f}$, ($f=t,b,c,\tau,\mu$), $h \rightarrow V V$, ($V = g, \gamma, W, Z$), and $h \rightarrow Z \gamma$. Additionally, in the LHCHCWG calculations, the Monte Carlo event generator {\tt ProPHECY4F}\,\cite{Bredenstein:2006rh,Bredenstein:2006nk,Bredenstein:2006ha} is also used to provide the partial widths for all possible four-fermion final states originating from $h \rightarrow ZZ$ and $h \rightarrow WW$ at both leading order (LO) and next-to-leading order (NLO).

The exotic decay of observed 125 GeV Higgs has been searched in various channels through pseudoscalars and yielded null results so far at the LHC. Searches have been performed
in the $4\mu$\,\cite{ATLAS:2018coo,CMS:2018jid},
4b\,\cite{ATLAS:2018pvw,ATLAS:2020ahi}, 2b2$\tau$\,\cite{CMS:2018zvv},
2$\mu$2$\tau$ or 4$\tau$\,\cite{ATLAS:2015unc,CMS:2017dmg,CMS:2018qvj,CMS:2019spf,CMS:2020ffa}
channels. However, since the pseudoscalar $A$ is quark-phobic and dominantly decays to the leptons, $4b$, $2b2\tau$ channels are not at all sensitive in the context of Type-X 2HDM. The $4\mu$ channel is also not sensitive compared to 2$\mu$2$\tau$ channel as the BR($A\to \mu\mu$) is suppressed by a factor $m_\mu^2/m_\tau^2$ compared to the BR($A\to \tau\tau$).
Thus, the most sensitive channel, in principle, becomes 4$\tau$. However, the $\tau$'s detected in their hadronic modes make the sensitivity of the search comparable to 2$\mu$2$\tau$ channel.
Importantly, all the prior searches have been carried out in the region where the pseudoscalar mass is less than 62.5 GeV, where it is produced on-shell.

Our primary attempt is to extend the search beyond the on-shell region, {\it i.e.} to consider the pseudoscalar mass greater than 62.5 GeV. In other words, whether or not the search for $h\to A A^*$, where $A^*$ represents the off-shell pseudoscalar, decay mode can be studied, and a potential signal can be observed. 
The primary obstacle to probing beyond the 62.5 GeV mass zone is the extremely low number of signal events (even at 100~fb$^{-1}$ of integrated luminosity) due to the presence of off-shell pseudoscalar and the overwhelming amount of background events. We, therefore, selected the dominant pseudoscalar decay mode through $\tau$-pairs to enhance the signal events considering HL-LHC in the future. The signal with $4\tau$'s has been tackled in the following way to suppress the background. 

The $\tau$ has two decay modes, {\it viz.} leptonic ($\approx 35\%$),
hadronic ($\approx 65\%$) decay modes. The $\tau$, via its hadronic
decay mode, is detected as a jet, called $\tau$-jet or $\tau_h$. The
$\tau$-jets  can be tagged with an efficiency of $\approx 60\%$ as
these are low multiplicity jets. When folded with the respective
branching ratios, one expects the $\tau$ can be identified at a rate
of $\approx 40\%$ in the hadronic mode, which is higher than its
leptonic decay modes. However, the main difficulty comes in managing
the SM QCD multijet and $t\bar t$ backgrounds for which one expects
billions of events in the HL-LHC. Although the misidentification
rate of a QCD jet tagged as a $\tau$-jet is as low as 1.0\%, due to
the huge number of events, the SM background remains present with a
significant amount. Thus, we look for the signals where two same-sign
$\tau$'s decay leptonically and two same-sign $\tau$'s decay
hadronically. This is possible because of the high efficiency in the
charge identification efficiency of $\tau$-jets. These efficiencies
are as high as 99\% and 70\% in the one-prong and three-prong decay,
respectively\,\cite{CMS:2022prd}. On the other hand, the reconstruction
efficiencies of the muons and electrons are very high, more than
97\%\,\cite{ATLAS:2016tmt,CMS:2023dsu}. For those reasons, we kept two
same-sign leptons and two same-sign $\tau$-jets as our main focus in
the final states.

The same search strategy can be applied to the on-shell region, where ATLAS and CMS already set upper limits on the product of the cross section of $pp\to h$ and BR($h\to A A$). In our study, we, however, probe for $\lamhAA$ coupling instead of the BR($h\to A A$) since we look into the off-shell region as well. In the off-shell region, the cross section of $pp\to h \to A A^* \to 4\tau$
has a dependency on the actual value of the decay width of $A$ through the Breit–Wigner. This effect is usually not present in on-shell decay mode, especially in the narrow width approximation, where Breit-Wigner contribution can be replaced by production cross section times the branching ratio. Therefore, the probe of $\lamhAA$ coupling will depend on the value of decay width, which is primarily controlled by the Yukawa coupling $y_{\tau\tau A}$. Hence, we expect our probe of $\lamhAA$ coupling to be less sensitive to $y_{\tau\tau A}$ in the on-shell case and more sensitive to $y_{\tau\tau A}$ in the off-shell case.

For our study, we have selected the pseudoscalar mass range between 40 GeV and 85 GeV.
We primarily divide this selected $m_A$ range into three regions, and we call them (a) on-shell region,
(b) intermediate region, and (c) off-shell region.
The first region is the on-shell region ($40 \leq m_A < 60$), where 125 GeV Higgs decays into
two on-shell pseudoscalars, followed by their decay to $\tau$'s. In
this region, the number of signal events is insensitive to the exact
value of $y_{\tau\tau A}$ coupling because the pseudoscalars are
produced on-shell and the $y_{\tau\tau A}$ coupling mainly contributes
to the BR($A \to \tau\tau$), which is almost 100\%. In the second
region ($60 \leq m_A \leq 65$), the 125 GeV starts decaying
through one on-shell and one off-shell pseudoscalars. Hence, the
corresponding event rate is highly sensitive to the decay width of the pseudoscalar.
Therefore, the number of signal events becomes sensitive to the value
of $y_{\tau\tau A}$ coupling. Lastly, $h$ decays mostly through one
on-shell and one off-shell pseudoscalar in the third region
($65 < m_A \leq 85$), with a little lower sensitivity to the
$y_{\tau\tau A}$ coupling than in the preceding regions.

To investigate the couplings in this study, we have taken a few
discrete pseudoscalar masses and performed the analysis at those
discrete mass points. Additionally, the cross section in the 4$\tau$
channel is sensitive to the actual value of the decay width of the
pseudoscalar $A$, especially in the intermediate and off-shell
regions. We thus take two different values of $y_{\tau\tau A}$ coupling
in the analysis. The mass points are chosen sparsely with a gap of
2.5~GeV in the on-shell region since the cross section is less
sensitive to $m_A$ in this region. However, in the intermediate region, the
cross section is sensitive to the exact mass values. Hence, in this
region, the mass points are chosen with a finer spacing of 0.5~GeV. In
the off-shell region, the mass points are chosen with a spacing of
2.5~GeV again.

\subsection{Signal and Background} \label{sec:sig-bkg}
We took the dominant channel (ggF) for 125 GeV Higgs production,
through which two pseudoscalars are produced, followed by decaying
into 4$\tau$'s, due to having almost 100\% branching ratio in the
di-tau mode. In addition, the cross section of the $h$ production is
significant in the one extra jet channel, where this extra jet
appears due to 
initial state radiation (ISR). So, the signal topology, in which we
are interested, is the production of $h$ plus up to one extra jet. The
process can be written as
\begin{align}
    pp &\rightarrow h + (j)\rightarrow A A^{(*)} + (j)\rightarrow 2\tau^\pm + 2\tau^\mp + (j),
\end{align}
where, by $A^{(*)}$, we mean on-shell $A$ or off-shell $A^*$ depending
on the region of our analysis.

The ($2l^{\pm} + 2\tau_h^{\mp}$) final state may arise from other
sources that resemble our intended signal process. Although vector
bosons predominantly contribute to the leptons and $\tau$'s
production, they can originate from other places also, like the
semileptonic decays of the $B$-hadrons. Furthermore, a major
challenge arises from the QCD jets, which can be misidentified as
$\tau$-jets, while the misidentification of muons and electrons is
insignificant. Considering all these, we list all potential sources of backgrounds
below.
\begin{align}
    pp &\rightarrow VV + (j), \\
    pp &\rightarrow t\bar{t}V, \\
    pp &\rightarrow t\bar{t}+ (j),
\end{align}
where $V$ represents vector bosons: $Z$, and $W^\pm$ bosons. In the
first process, four $\tau$'s can directly arise from the decays of the
two vector bosons and can be regarded as irreducible backgrounds if
both of them are $Z$ bosons. In scenarios involving one $Z$ boson and
one $W$ boson, leptons can arise through $Z$ boson, while one $\tau_h$
originates from $W$ boson, and the extra QCD jet could be
misidentified as a $\tau$-jet. This background contributes very little
to the actual signal topology since two leptons are oppositely charged
and the same-sign dilepton pair appears only due to the
misidentification of the charge of one of the two leptons. The
background for two $W$ bosons is negligible because it requires a QCD
jet to be misidentified as a lepton, and the other lepton and one
$\tau_h$ can come from $W$ bosons, whereas second $\tau_h$ would,
again, have to come from a misidentified QCD jet. In the background process $pp \rightarrow t\bar{t}V$, the top quark
has 100\% decay into a bottom quark and a $W$ boson, leaving
$b\bar{b}W^+W^-V$ as the hard process. This process has various
combinations for misidentifying leptons and $\tau_h$'s. For instance,
in the case of $V=W$, if two out of three $W$ bosons undergo hadronic
decay, one lepton can arise from the remaining $W$ boson while another
may originate via semileptonic decay of $B$-hadron. Similarly, one
$\tau_h$ can also be produced via $B$-hadron when the QCD jet is
mistakenly detected as another $\tau_h$. Likewise, the third process
$pp \rightarrow tt+(j)$ yields $b\bar{b}W^+W^-+(j)$ as the hard
process. Further, if one $W$ boson decays hadronically, two leptons
can arise from one $W$ boson and one $B$-hadron. The other $B$-hadron
could generate one $\tau_h$, while the other $\tau_h$ might come as a
result of misidentification of a QCD jet.

For the event generation, we have used the Mathematica-based package
{\tt SARAH}\,\cite{Staub:2013tta} to generate the UFO model file. The
spectrum files have been generated from the
{\tt SPheno}\,\cite{Porod:2011nf} package, which calculates the Higgs
masses, all types of couplings, decay widths, branching ratios, etc.
All the signals and backgrounds have been simulated at the parton
level at a centre-of-mass energy of 14 TeV using the well-known
package, {\tt MadGraph5\_aMC@NLO}\,\cite{Alwall:2014hca}. Further
showering and hadronization have been done through
{\tt PYTHIA8}\,\cite{Bierlich:2022pfr} followed by
{\tt Delphes}\,\cite{deFavereau:2013fsa} for detector simulation. Jets
are clustered using the anti-$k_t$ algorithm\,\cite{Cacciari:2008gp}
with a radius of 0.5 GeV. As we have an additional jet, which has a chance
to get generated at the hard process level as well as a result of
parton showering, in the final state, we have used the standard
practice of Matching and Merging\,\cite{Frederix:2012ps} through
{\tt MadGraph5\_aMC@NLO} and {\tt PYTHIA8} to eliminate the double-counting of the
contribution of the events in the common phase space. All the
background events have been generated at the leading order (LO).
We have then used corresponding $k$-factors of the next-to-leading
order (NLO) to scale the cross sections. The $k$-factors are 1.38,
1.57, 1.60, and 2.01, 1.72 for $t\bar t Z$\,\cite{Kardos:2011na},
$t\bar t W$\,\cite{vonBuddenbrock:2020ter},
$t\bar t$\,\cite{Czakon:2013goa},
$\tau^+\tau^- W + \text{jets}$\,\cite{Grazzini:2016swo}, and
$2\tau^\pm + 2\tau^\mp + \text{jets}$\,\cite{Cascioli:2014yka},
respectively.
\begin{figure}[!h]
\begin{center}
\includegraphics[width=0.7\textwidth]{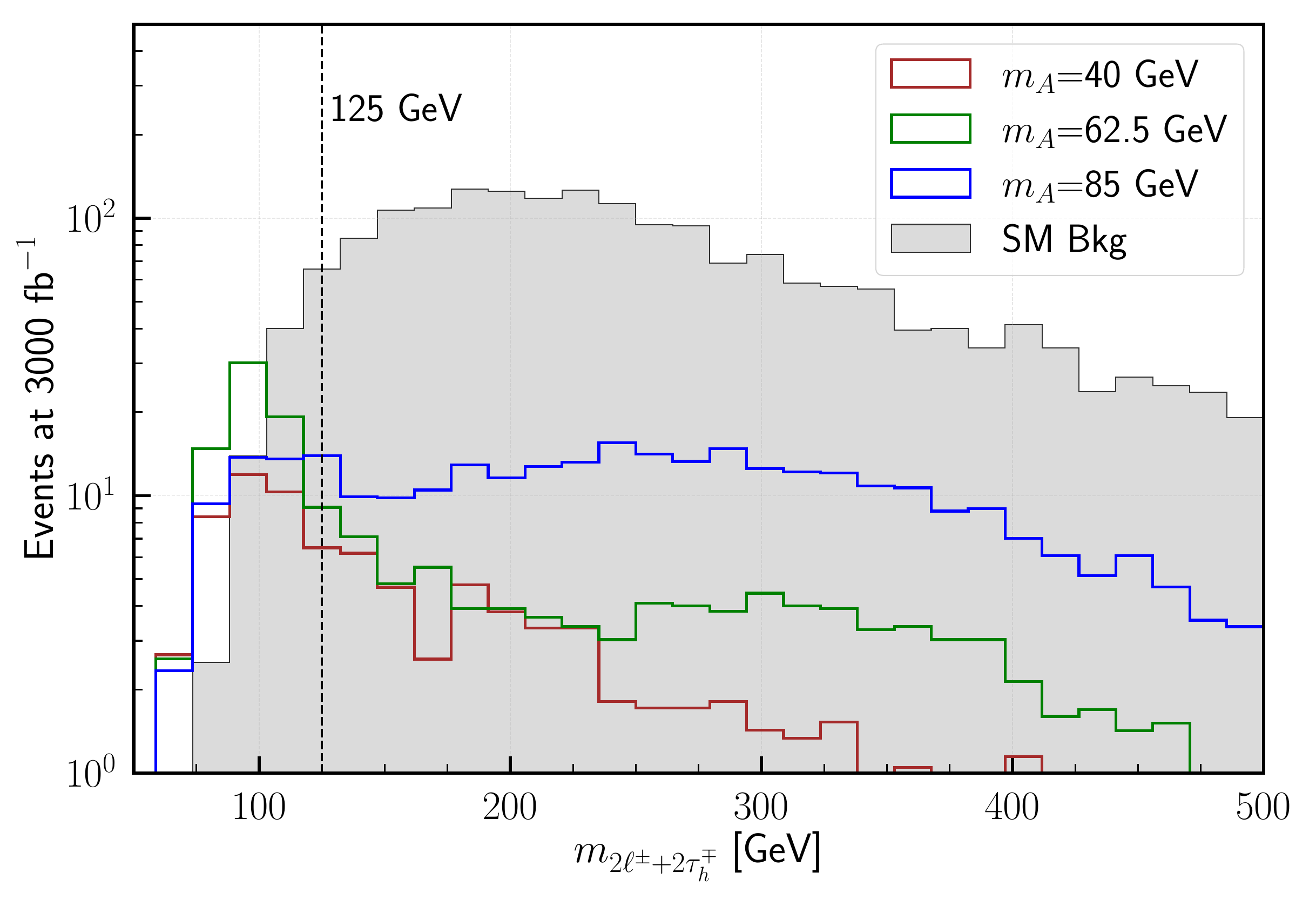}
\end{center}
\caption[]{Distribution of the invariant mass of
(2$\ell^\pm$ + 2$\tau_h^\mp$) for background (shaded) and signal for
different pseudoscalar masses $m_A$ = 40 GeV (blue), 62.5 GeV (red),
and 85 GeV (green). The parameters corresponding to the three pseudoscalar masses are given in Table~\ref{table:parameter}.}
\label{fig:inv_mass}
\end{figure}

Regarding the distinctive features of the signal process, the key
characteristic is all $\tau$'s are coming from 125 GeV Higgs, so at
the parton level, their invariant mass would sharply peak around the
Higgs mass. Furthermore, due to the missing neutrinos from the decays
of $\tau$'s, the invariant mass tends to shift towards lower values
as the final state contains (2$\ell^\pm$ + 2$\tau_h^\mp$) with six
missing neutrinos. The invariant mass
($m_{2\ell^\pm + 2\tau_h^\mp}$) distribution of these four
objects is shown in Fig.~\ref{fig:inv_mass} for three different $m_A$
values and the corresponding parameters are given in Table~\ref{table:parameter}.

\begin{table}[h!]
\centering 
\begin{tabular}
{c@{\hspace{1cm}}c@{\hspace{1cm}}c@{\hspace{1cm}}c@{\hspace{1cm}}}
\hline
\toprule
 & \multicolumn{3}{c}{\textbf{Benchmark Points}} \\ [-10pt]
\textbf{Parameters} & & & \\ [-10pt]
 & \textbf{$m_A$ = 40} & \textbf{$m_A$ = 62.5} & \textbf{$m_A$ = 85} \\
\hline
\cmidrule(lr){2-4}
\midrule
$\lambda_1$ &  0.09   &  0.10   &  0.27   \\
$\lambda_2$ &  0.13   &  0.13   &  0.13   \\
$\lambda_3$ &  4.60   &  3.73   &  2.37   \\
$\lambda_4$ & -2.27   & -1.90   & -1.44   \\
$\lambda_5$ &  2.32   &  1.84   &  1.12   \\
$m_{12}$ (GeV)   & 41.44   & 37.26   & 30.01   \\
$\tan\beta$ & 83.10   & 83.10   & 83.10   \\
\hline
$m_A$ (GeV) & \textbf{40.00} & \textbf{62.50} & \textbf{85.00} \\
$\lamhAA$ (GeV)  &  \textbf{0.27}   &  \textbf{0.91}   & \textbf{45.99}   \\
$y_{\tau\tau A}$&\textbf{0.60} &  \textbf{0.60}   &  \textbf{0.60}   \\
\hline
\end{tabular}

\caption{The parameter points for different pseudoscalar masses used in Fig~\ref{fig:inv_mass} to discriminate signal and background. The parameters $m_A$, $\lamhAA$ and $y_{\tau\tau A}$ are dependent parameters.}
\label{table:parameter}
\end{table}

Depending on the $\lamhAA$ and $y_{\tau \tau A}$ couplings,
the peak height would be different.
Note that there is another wider bump that appears around 300 GeV in
the signals, which come from the presence of other couplings that
produce the same final states, {\it i.e.} $4\tau + (j)$ but without
$h$ being present in the $s$-channel resonance. On the other hand, the
backgrounds, especially the dominant $ZZ$ and $t\bar{t}$ backgrounds,
tend to have the four-object invariant mass at a higher ($>150$~GeV)
value.
Therefore, in the final selection, we used an upper cut on
$m_{2\ell^\pm + 2\tau_h^\mp}$, which suppresses the
backgrounds and eliminates contributions from diagrams without $h$ in
the $s$-channel. To discriminate signals from the backgrounds, we used
the following important cuts:
\vspace{-2mm}
\begin{equation}
\vspace{-1mm}
    \begin{aligned}
        p_T^{j_1} &\geq 25\ {\rm GeV}, \quad
        p_T^{\ell_1} \geq 15\ {\rm GeV}, \quad
        m_{2\ell^\pm + 2\tau^\mp} \leq 100\ \text{GeV},
    \end{aligned}
\end{equation}
where $p_T^{j_1}$ and $p_T^{\ell_1}$ are the transverse momentum of
the leading jet and the leading lepton.

\subsection{Results and Discussions} \label{sec:result-dicuss}
\begin{figure}
\subfloat[]{\includegraphics[width=0.5\textwidth]{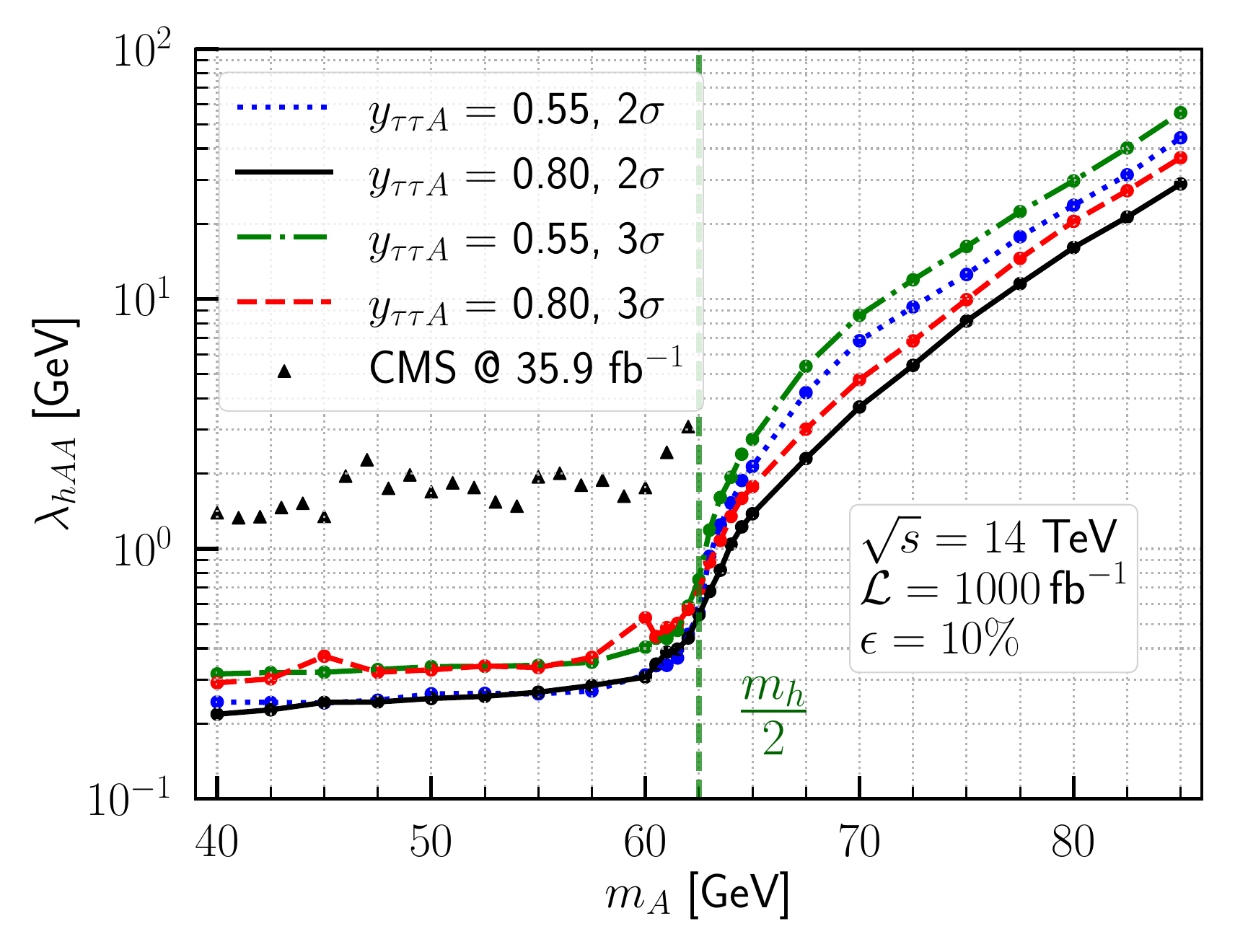}\label{fig:lum_1000}}\hfill
\subfloat[]{\includegraphics[width=0.5\textwidth]{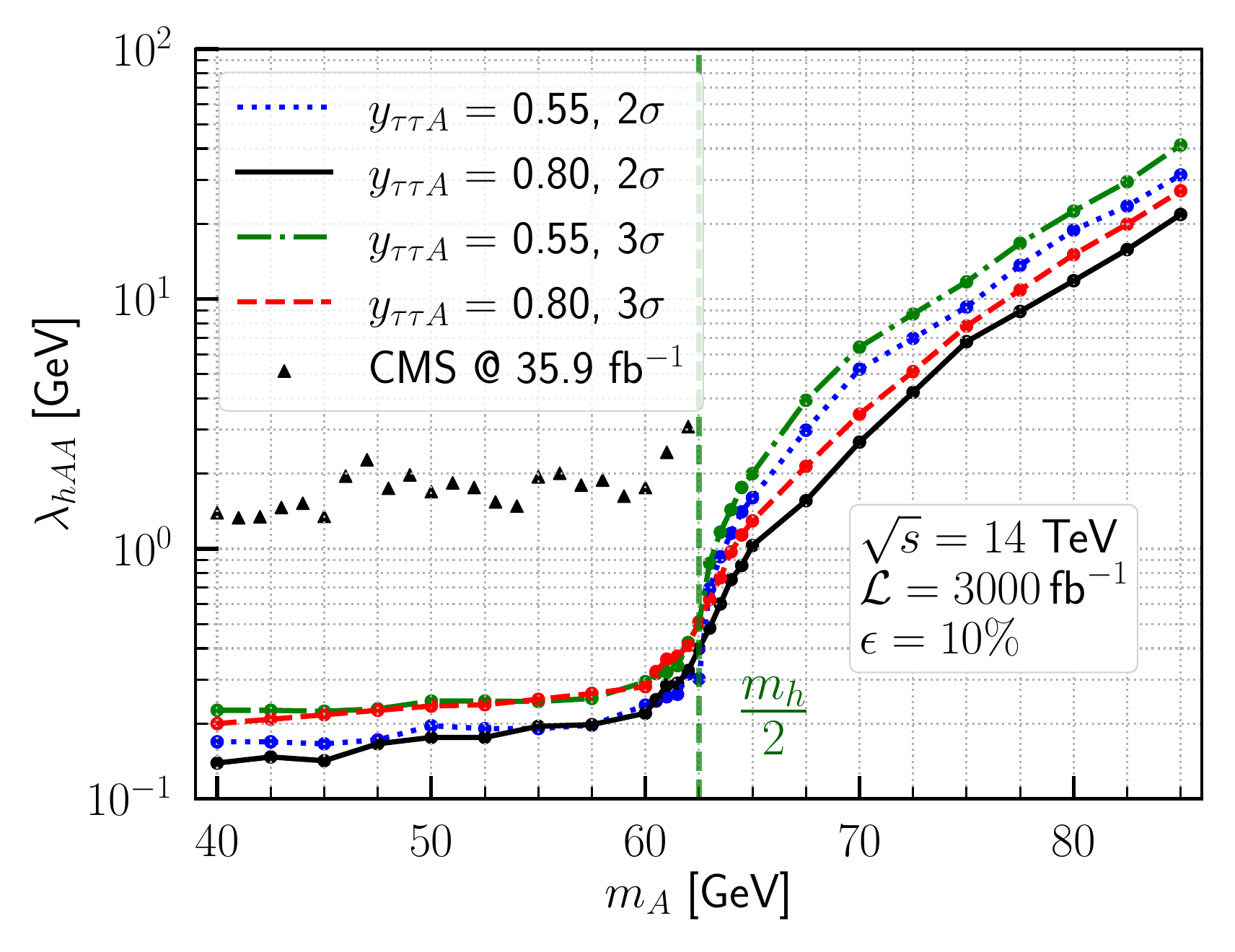}\label{fig:lum_3000}}
\captionsetup{justification=raggedright,singlelinecheck=on}
\caption{Discovery limit of $\lamhAA$ coupling as a function of $m_A$
at $2\sigma$ and $3\sigma$ 
at different values of the dimensionless coupling $y_{\tau\tau A}$.
The limits have been calculated with integrated luminosity (a) 1000
fb$^{-1}$, and (b) 3000 fb$^{-1}$ with a moderate 10\% systematic
uncertainty at a centre-of-mass energy of 14 $\text{TeV}$ LHC.
The black triangle points represent the upper bound on the $\lamhAA$
coupling in Type-X 2HDM, determined from observed data at
CMS\,\cite{CMS:2018qvj} with a 95\% confidence level at an integrated
luminosity of 35.9 fb$^{-1}$ at 13 $\text{TeV}$ LHC.}
\label{fig:main_plot}
\end{figure}

To start with the findings, we concentrate our efforts on the three areas mentioned.
That is to say, we searched the collider signature of the pseudoscalar in the
intermediate and off-shell region, where the pseudoscalar mass is
larger than half of $m_h$ ($m_A \geq \frac{m_h}{2}$). We
further investigated whether our approaches yield better results on the
BR($h \rightarrow AA$) than the earlier searches at the LHC\,\cite{CMS:2018qvj} in the
region where pseudoscalar mass is less than half of the mass of $h$.
For this, we first calculate the significance $\mathfrak{S}$ for a
given value of $\lamhAA$. Given the number of signal events $S$ and
number of background events $B$ after the final selection cut, one
usually calculates the significance of discovering the signal. In
the presence of relative systematics uncertainty $\epsilon$, the
significance can be calculated using the following formula.
\begin{eqnarray}
\mathfrak{S} 
  =\sqrt{2}\left[(S+B)\ln\left(1+\frac{S}{B+\epsilon^2 B(S+B)}\right)
  -\epsilon^{-2}\ln\left(1+\frac{\epsilon^2 S}{1+\epsilon^2 B}\right)
   \right]^{\frac{1}{2}},
\end{eqnarray}
We then provide the discovery limit of potentially observing the
signal at two specific values of signal significance, {\it viz.}
2$\sigma$ and 3$\sigma$ in
Fig.~\ref{fig:main_plot}. The blue dotted and black solid lines
represent the values of $\lamhAA$ for which the signal could be
observed at 2$\sigma$ for $y_{\tau\tau A} = 0.55$ and 0.80,
respectively. The green dash-dotted and red dashed lines represent
the curves at 3$\sigma$ for the above two values of $y_{\tau\tau A}$.
The two panels, {\it i.e.} Figures~\ref{fig:lum_1000} and
\ref{fig:lum_3000} correspond to the discovery limits of $\lamhAA$
couplings assuming 1000 fb$^{-1}$ and 3000 fb$^{-1}$ luminosity,
respectively. In all the curves, signal significance is calculated
considering a moderate 10\% systematics uncertainty.
The essential features of the discovery limits can be noted down as
follows
\begin{itemize}
  \item The numerical value of the discovery limits of $\lamhAA$ at
  3$\sigma$ is larger than that at 2$\sigma$ since the former sets a higher
  level of signal significance and hence can only probe large values.
  \item In the on-shell ($m_A < m_h/2$) region, the limit is almost
  insensitive to the actual value of $y_{\tau\tau A}$ since both
  pseudoscalars are produced on-shell, and then each decays into a
  pair of $\tau$'s. The branching ratio of $A \to \tau\tau$ is almost
  100\% and does not depend on the exact value of $y_{\tau\tau A}$ in
  this parameter region.
  \item In the off-shell ($m_A > m_h/2$) region, where one
  pseudoscalar is produced off-shell, the limit is dependent on the
  value of $y_{\tau\tau A}$. This is because the larger value of
  $y_{\tau\tau A}$ coupling makes the decay width of the pseudoscalar
  larger, which contributes to higher di-$\tau$ production from an
  off-shell $A$. Higher production implies that the signal can be
  probed deeper, and hence, the discovery limit can be set to a smaller
  value. This can be seen in Figs.~\ref{fig:lum_1000} and
  \ref{fig:lum_3000} where the two curves corresponding to
  $y_{\tau\tau A} = 0.80$ sets stronger limits than the two lines
  corresponding to $y_{\tau\tau A} = 0.55$.
  \item Finally, higher integrated luminosity means being able to
  probe smaller values of $\lamhAA$. This feature is demonstrated in
  the figure, where one can see that each curve in
  Fig.~\ref{fig:lum_3000} is shifted downward below with respect to
  the corresponding curve in Fig.~\ref{fig:lum_1000}. However, the
  systematics, which has been taken to be 10\% in this analysis,
  tends to reduce the difference between the two cases. 
\end{itemize}

Since no pseudoscalar search for masses greater than 62.5 GeV is
reported in the available literature, our proposed methods of finding
such signals in the two same-sign dilepton and two same-sign
$\tau$-jets will help in the probe of $\lamhAA$ couplings in the
HL-LHC. On the other hand, the searches of $h$ decaying to a pair of
pseudoscalars have been performed in the region where the
pseudoscalar mass is less than half of the mass of $h$ at CMS in 2017
at 19.7~fb$^{-1}$ of integrated luminosity\,\cite{CMS:2017dmg}.
Further, a follow-up search in 2018 at 35.9 fb$^{-1}$ integrated
luminosity improved the limit on the branching ratio by more than a
factor of two\,\cite{CMS:2018qvj}. In these searches, an upper limit
on the branching ratio of $h\to AA$ has been set at 95\% C.L.~in the
$2\mu+2\tau_h$ channels, considering one pseudoscalar to decay to two
$\mu$'s and the other to decay to two $\tau$'s. Hence, in these
searches, both the $\mu$ pair are required to have opposite signs, and
the same for the $\tau_h$ pair. The ratio of BRs in the $\mu$ mode to
$\tau$ mode is assumed to be $m_\mu^2/m_\tau^2$. Assuming this ratio,
one can get the upper limit on BR$(h\to AA)$. We then used the
Eqs.~(\ref{eq:brhAA}) and (\ref{eq:dwhAA}) to obtain a limit on
$\lamhAA$. The black triangle points in Fig.~\ref{fig:main_plot}
represent the current upper limit on $\lamhAA$ calculated from the
upper limit on the BR($h \to AA$) provided by the previous search in
CMS at 35.9 fb$^{-1}$\,\cite{CMS:2018qvj}.

One might ask how our proposed analyses, which are performed assuming 1000~fb$^{-1}$ or 3000~fb$^{-1}$ luminosity, fair against the CMS analysis at 35.9~fb$^{-1}$. We explain this in the following way.
At integrated luminosity $\mathcal{L}$, the signal significance for $h\to A A$ is
proportional\footnote{This relation also assumes the denominator of
the Eq.~(\ref{eq:brhAA}) changes negligibly since the 95\% C.L.~limit
by CMS on the BR($h\to AA$) is less than a few per cent.} to
$\lamhAA^2\times\sqrt{\mathcal{L}}$ if we consider no systematics and
the naive $S/\sqrt{B}$ formula. Now, if $\lamhAA$ becomes the
observed limit at luminosity $\mathcal{L}_0$, one might simply
estimate that the required luminosity to probe a coupling
$\frac{\lamhAA}{\alpha}$ ($\alpha$ being a scale factor) would be
$\alpha^4 \mathcal{L}_0$. Observing that the projected discovery limit
of $\lamhAA\approx 0.1$~GeV at 3000~fb$^{-1}$ in our proposed channel
is almost an order of magnitude better than the CMS result at 35.9
fb$^{-1}$, the required luminosity to probe $\lamhAA\approx 0.1$~GeV
in the CMS channel would require almost $35.9\times10^4$ fb$^{-1}$ luminosity.
Thus, our proposed method provides an improved limit in the region
$m_A \leq \frac{m_h}{2}$ considering 10\% systematic uncertainty.

The potential scope for quality improvement in our analysis can be performed in two places. Firstly, the kinetic distributions, such as the invariant
mass of $2\ell^\pm + 2\tau^\mp$, are generated at the LO, and only the
overall cross section is scaled by NLO $k$-factors. This can be
improved to the kinetic distribution at NLO. Secondly, we have used
the simplified $\tau$-identification as implemented in {\tt Delphes}
which can be refined to more realistic $\tau$ taggers. The {\tt Delphes} $\tau$-tagger uses an efficiency-based implementation, i.e. an efficiency is used in the $\tau$-objects, and a misidentification rate is used for other objects to incorporate the effects of faking other objects as $\tau$-jets. In our analysis through {\tt Delphes}, we have chosen a working point with efficiency as 60\% and jet$\to$tau misidentification probability as 1\%\,\cite{CMS:2018jrd}. With the latest ML-based {\tt DeepTau} tagger\,\cite{CMS:2022prd}, the misidentification rate can be reduced by a factor of two, especially for the jets coming from $t\bar t$ sample. Therefore, a factor of four reduction in the dominant $t\bar t$ background is possible, allowing for a two-fold improvement in the significance. Although a thorough analysis is required to quote a concrete number, a significant improvement is expected with the help of the state-of-the-art {\tt DeepTau} tagger.

\section{Summary and Conclusion} \label{sec:summ}
The exotic decay modes of $h$ into two on-shell pseduscalars ($A$) is a primary signature of a Type-X 2HDM whenever
the decay is kinematically allowed. The CMS and ATLAS collaborations
at the LHC have met with null results in these efforts in the search
for $h\to AA$ mostly in the $4\tau$ channel, for $m_A < 62.5$ GeV. 

In this work, we have investigated, in the context of the collider signatures of $h\to AA^{(*)}$ in $4\tau$ channel. We show that it is possible to achieve visibility for one off-shell pseudoscalar if we look for the final state comprising two same-sign leptons along with two same-sign $\tau$-jets.
In this channel, the SM background, as we show, can be strongly eliminated.
The important feature of our proposal is that the probe is feasible for $m_A > m_h/2$, where one pseudoscalar is produced off-shell. 
The minimum $hAA^*$ interaction strength that can be thus probed at the high-luminosity run of the LHC has been obtained for $m_A$ all the way to 85 GeV.
We present our results in terms of the potential discovery limit of the $\lamhAA$ coupling at the $2\sigma$ and $3\sigma$.
We further have shown the potential improvement in the probe of $hAA$ coupling for the pseudoscalar masses $m_A < m_h/2$ at the LHC with 1000~fb$^{-1}$ and 3000~fb$^{-1}$ integrated luminosity. To achieve the probe of $\lamhAA$ received with our strategy, the conventional searches would require approximately $3.6\times 10^5$ fb$^{-1}$ integrated luminosity at a 14 TeV hadron collider.

\section*{Acknowledgements}
The authors are thankful to the High Performance Computing (HPC) Cluster (Kepler) facility provided by the Department of Physical Sciences at IISER Kolkata. S.~S.~thanks to the Council of Scientific and Industrial Research (CSIR) for funding this project.

\providecommand{\href}[2]{#2}\begingroup\raggedright\endgroup

\end{document}